\documentclass[a4paper,twocolumn,accepted=2022-04-19]{quantumarticle}
\pdfoutput=1
\usepackage[utf8]{inputenc}
\usepackage[english]{babel}
\usepackage[T1]{fontenc}
\usepackage{amsmath}
\usepackage{hyperref}
\usepackage{blindtext}
\usepackage{bm}
\usepackage{braket}
\usepackage{amsmath}
\usepackage{amssymb}
\usepackage{graphicx}
\usepackage{tikz}
\usepackage{lipsum}
\usepackage{braket}
\usepackage{cite}

\begin{document}

\title{Back to sources---the role of losses and coherence in super-resolution imaging revisited}

\author{Stanisław Kurdziałek}
\affiliation{Centre for Quantum Optical Technologies, Centre of New Technologies, University of Warsaw, Banacha 2c, 02-097 Warszawa, Poland}
\affiliation{Faculty of Physics, University of Warsaw, Pasteura 5, 02-093 Warszawa, Poland}
\orcid{0000-0001-8877-1421}
\email{s.kurdzialek@student.uw.edu.pl}
\maketitle

\begin{abstract}
  Photon losses are intrinsic for any translationally invariant optical imaging system with a non-trivial Point Spread Function, and the relation between the transmission factor and the coherence properties of an imaged object is universal---we demonstrate the rigorous proof of this statement, based on the principles of quantum mechanics. The fundamental limit on the precision of estimating separation between two partially coherent sources is then derived. The careful study of the role of photon losses allows to resolve conflicting claims present in previous works. We compute the Quantum Fisher Information for the generic model of optical 4f imaging system, and use prior considerations to validate the result for a general, translationally invariant imaging apparatus.  We prove that the spatial-mode demultiplexing (SPADE) measurement, optimal for non-coherent sources, remains optimal for an arbitrary degree of coherence. Moreover, we show that some approximations, omnipresent in theoretical works about optical imaging, inevitably lead to unphysical, zero-transmission models, resulting in misleading claims regarding fundamental resolution limits.
\end{abstract}

\section{Introduction} 

The rapid development of super-resolution optical imaging prompted physicists to rethink the way in which the performance of different imaging methods is quantified. The most promising approach is based on (quantum) estimation theory \cite{bettens1999model,Ram2006rohtua,tsang,tsang2019resolving,lupo2016ultimate,nair2016far}, where the attainable estimation precision of object parameters is used to assess the quality of a given imaging scheme. In the simplest and most commonly examined scenario, imaging of two equally bright point sources is considered. For a given measurement, the variance of any locally unbiased estimator of the separation ($s$) between the sources is lower-bounded by the inverse of Fisher Information (FI). When $s$ is smaller than the width of the system Point Spread Function (PSF), it is hard to distinguish the images of two sources, which is reflected by the FI vanishing for $s \rightarrow 0$. This fact, true for standard measurement of the light intensity in the image plane, is known as the \textit{Rayleigh curse}. Surprisingly, the Rayleigh curse can be omitted, if the measurement of light in the image plane can be freely chosen \cite{tsang}. Quantum Fisher Information (QFI) associated with the estimation of $s$, which is the classical FI maximized over all possible measurements, does not depend on $s$, and does not vanish for $s \rightarrow 0$. The above statement is valid for an idealized case, in which no noise apart from shot noise is present, and $s$ is the only unknown parameter characterizing the object, which consists of exactly two sources whose centroid is known \textit{a priori} \cite{tsang}. More realistic and complex scenarios, involving different types of noise \cite{tsang2019resolving,len2020resolution, oh2020quantum}, more complicated objects \cite{Zhou2019,Albarelli2020, tsang2019semiparametric}, and different types of light sources \cite{lupo2016ultimate, nair2016far}, have also been examined. All these extensions are based on a formalism from the seminal work \cite{tsang}---it is, therefore, crucial to understand all the aspects of the simplified model presented there.

One of the aspects, which in our opinion, has not been given enough attention so far, is the role of photon losses, which are present for all passive, linear, translationally invariant imaging systems with non-trivial PSFs. As we will show in Section~\ref{sec:losses}, this statement can be derived from the fundamental principles of quantum mechanics, without referring to any specific imaging apparatus---moreover, the relation between the input field, the system PSF, and the loss ratio is universal. Consequently, it is necessary to take into account photon losses even in most basic scenarios, in which all imperfections present in experiments are neglected---lossless models are not only unrealistic, but also unphysical and logically inconsistent. 

The impact of fundamental losses caused by the imaging system is particularly important, when the observed object is (at least partially) coherent. In such a case, the total power of the signal observed in the image carries the information about the object shape because the photon loss ratio depends on the distribution of the input coherent field. Therefore, paradoxically, the presence of losses may improve the obtainable estimation precision of some of the object's parameters. This observation, together with the rigorous study of the losses characteristic shown in Section \ref{sec:losses}, will be used in Section \ref{sec:coherence} to resolve recent disputes concerning the influence of partial coherence between two sources on the maximal precision of the estimation of their separation \cite{larson2018resurgence, tsang2019resurgence,larson2019resurgence, hradil2019quantum, lee2019surpassing, tsang2019resolving, liang2020coherence, wadood2021superresolution}. Contradictory claims, that have appeared in this debate, were caused by the fact, that some authors use unphysical, lossless models of the imaging apparatus.

Even though non-unit transmission factor $T$ is present in most of the far-field imaging models \cite{tsang, nair2016far, lupo2016ultimate}, authors tend to neglect the fact that their simplifying assumptions may lead to the necessity of assuming that $ T \rightarrow 0$  in order for their models to be physical. For example, translational invariance of a system implies that $T=0$ for objects consisting of a finite number of point sources, and also for all fully incoherent objects (see Section \ref{sec:point}). This issue may be fixed, for example, by replacing point sources with more realistic emitters, each of which is described by some narrow, but not point, distribution of coherent field. As will be demonstrated in Section \ref{sec:point}, this more physical model leads to new fundamental limits associated with a single light source localization.
\section{Losses in imaging systems}
\label{sec:losses}

Before we provide a fundamental, quantum mechanics-based description of photon losses, let us first assume that the object plane of our system is described by the classical distribution of the electric field perpendicular to the optical axis, $E(\bm r)$ ($E$ is a complex scalar because it contains phase information, and polarization is neglected). For linear, translationally invariant, diffraction-limited imaging system, the field distribution observed in the image plane is $ (E \ast H)(\bm r)$, where~$\ast$ denotes the convolution, $|H(\bm r)|^2$ is a PSF, and $H(\bm r)$ itself is called the coherent Point Spread Function (cPSF) \cite{barrett2013foundations}. The Fourier transform of the initial field distribution, 
\begin{equation}
\label{fourier}
\hat E (\bm k) \equiv \int E(\bm r) e^{-i 2 \pi \bm k \bm r} \textrm{d} \bm{r} ,
\end{equation}
becomes $\hat E (\bm k) \cdot \hat H (\bm k)$ after imaging system transformation. That means, that the signal power associated with different spatial frequencies must be modified when the resulting PSF is wider than the Dirac delta. For a passive system, no spatial frequency of $ E (\bm r)$ can be strengthened ($|\hat H (\bm k)|\le 1$), and therefore the described modification can be achieved by photon losses only. 

\begin{figure}[t]
\includegraphics[width=1. \linewidth]{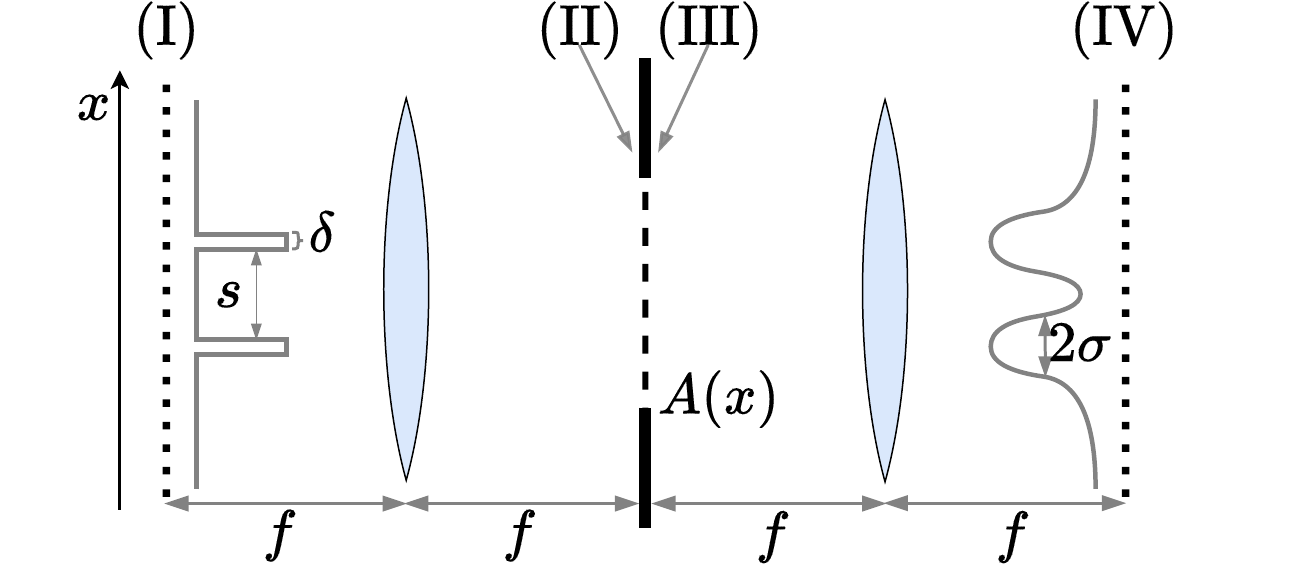}
\caption{The diagram shows the $4f$ imaging system, a well-known model of e.g. optical microscope. Lenses are ideal, so the complex field  amplitude in the right focal plane of each lens is the Fourier transform of the field in its left focal plane [light propagates from the object plane (I) to the image plane (IV)]. All system imperfections are modeled by the aperture with transmission coefficient $A(x)$, which is the source of photon losses and image blurring. The binary object discussed in Sec. \ref{sec:coherence} and its image are schematically drawn with a gray line. }
\label{4f}
\end{figure}

The above reasoning can be illustrated and make more quantitative by analyzing a so-called \textit{4f} system \cite{barrett2013foundations}. Such a system consists of two lenses with equal focal lengths ($f$), the setup geometry is shown in Fig.~\ref{4f}.

From now on, 1D imaging is considered, which is enough to show the crucial aspects of the studied problem. The aperture with transmission coefficient $A(x)$, placed between the two lenses, is the only source of image imperfections, and determines both the shape of the PSF and photon losses ratio. 
The lenses are assumed to be ideal, as their finite size can be modeled by a proper change of $A(x)$. The magnification of the considered system is 1, but all results are valid for any magnification, which is generally the ratio of the focal lengths of two lenses. 

The field amplitude distribution in the plane just before the aperture (II) is a Fourier transform of the field in the object plane \cite{barrett2013foundations}:
\begin{equation}
 E_\textrm{II}(x) = \frac{1}{\sqrt{f \lambda}} \hat{E}_\textrm{I} \left( \frac{x}{f \lambda} \right),  
\end{equation}
where $\lambda$ is the light wavelength, $\hat f(k)$ generally denotes the Fourier transform of $f(x)$ (convention from \eqref{fourier} is adopted for a 1D case).
From now on, let us assume that $f \lambda = 1$, which is equivalent to choosing $\sqrt{f \lambda}$ as the length unit.
The aperture modifies the field such that it becomes $E_\textrm{III}(x) = E_\textrm{II}(x) \cdot A(x)$ in plane (III). The ratio of the signal power transmitted through the aperture, which is also the probability of photon transmission, is equal to
\begin{equation}
\label{pphi1}
    T = \frac{\int_{- \infty}^{\infty} |E_\textrm{II}(x) A(x)|^2 \textrm{d} x}{\int_{- \infty}^{\infty} |E_\textrm{II}(x)|^2 \textrm{d}x},
\end{equation}
 The field observed in the image plane (IV), which is a Fourier transformed $E_\textrm{III}$, is 
 \begin{equation}
     E_\textrm{IV}(x) =  \left( E_\textrm{I}^{(-)} \ast \hat A \right)(x),
 \end{equation}
 where $E_\textrm{I}^{(-)}(x) = E_\textrm{I}(-x)$.
The above reasoning shows that $ \hat A \left(x\right)$is the system cPSF.  
 
In order to describe the image formation process using a quantum terminology, one should interpret the normalized distribution of the electric field $E(x)$ as a photon wave function $\psi(x)$ (as in e.g. \cite{hradil2019quantum}).
 Such a description is complete when no correlations between subsequent photons are present, which will be assumed in this paper. Let us consider the time of observation in which the object emits a photon with a probability $p_\textrm{em} \ll 1$, and the probability of emission of two or more photons is negligible. The quantum state of light emitted by a fully coherent object placed in plane (I) is then
 \begin{equation}
 \label{eq5}
     \rho_\textrm{I} = p_\textrm{em} \ket{\psi_\textrm{in}} \bra{\psi_\textrm{in}} + \left(1- p_\textrm{em} \right) \ket{0} \bra{0},
 \end{equation}
 where $\ket{0}$ is a zero-photon state, and $\ket{\psi_\textrm{in}}$ can be represented by a one-photon wave function in a position basis:
 \begin{equation}
 \label{psiin}
     \braket{x|\psi_\textrm{in}} = \psi_\textrm{in}(x) = \mathcal{N}_1 E_\textrm{I}(x),
 \end{equation}
 where $\mathcal{N}_1$ is a normalization factor.
 The transformation of the quantum state between planes (I) and (IV) is linear, state $\ket{0}$ is transformed into $\ket{0}$ (no photons are created), and a one-photon part becomes
\begin{equation}
\label{sout1}
    \ket{\psi_\textrm{in}} \bra{\psi_\textrm{in}} \longrightarrow T \ket{\phi_\textrm{out}} \bra{\phi_\textrm{out}} + (1-T) \ket{0} \bra{0},
\end{equation}
where $T$ is given by \eqref{pphi1}, and $\ket{\phi_\textrm{out}}$ is a normalized one-photon state characterized by wave function
\begin{equation}
\label{sout2}
 \phi_\textrm{out}(x) = \mathcal{N}_2 E_\textrm{IV}(x) = \mathcal{N}_2 \mathcal{N}_1^{-1}  \left(\psi_\textrm{in}^{(-)} \ast \hat A\right)(x)   ,
\end{equation}
where $\mathcal{N}_2$ is a normalization factor, generally different than $\mathcal{N}_1$, $\psi_\textrm{in}^{(-)}(x) =\psi_\textrm{in}(-x)$. 
In order to fully describe an output state in terms of an input state and the shape of the aperture, one should supplement \eqref{sout2} with the expression for transmission,
\begin{equation}
\label{Tq}
    T = \int_{-\infty}^\infty \left| \hat \psi_\textrm{in}(x) A(x) \right|^2 \textrm{d}x,
\end{equation}
which can be obtained from \eqref{pphi1}.
Equations (\ref{sout2}) and (\ref{Tq}) constitute the explicit relation between the input field, the shape of the cPSF, 
 and the system transmission $T$.
 
 It is not clear at this point, how universal this relationship is, especially as it leads to a non-intuitive conclusion, that $T=0$ for a single point source in the image plane for any reasonable cPSF (this will be further discussed in Section \ref{sec:point}).
  \begin{figure*}[t]
    \centering
    \includegraphics[width=1.\textwidth]{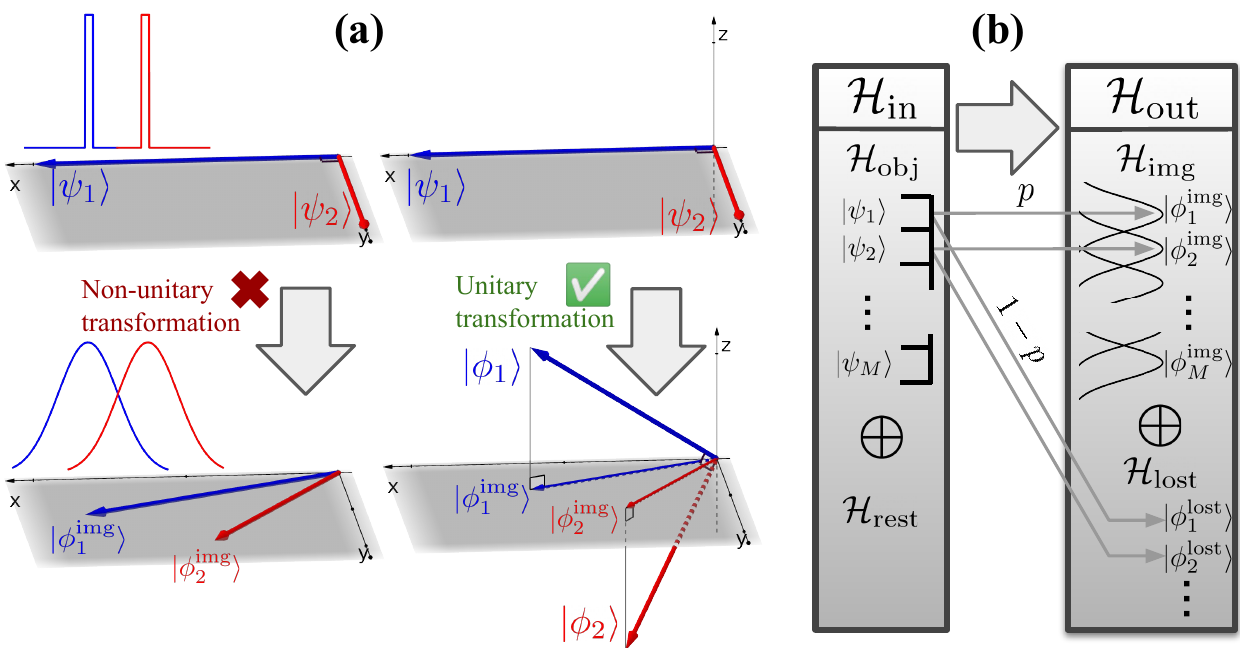}
    \caption{A unitary transformation preserves the scalar product---when the images of the two orthogonal modes of the object plane overlap, losses must be present. The basic intuition behind this argument is shown in part (a). In this work, a more general scenario, with many possible input modes, is quantitatively studied. The scheme of this general scenario is shown in (b).}
    \label{fig:losses}
 \end{figure*}
 Let us therefore go beyond the 4f imaging system case, and examine a general passive, linear, translationally invariant, optical imaging system, whose input modes annihilation operators are  $\hat a_1, \hat a_2, ..., \hat a_N$. Some of these modes ($\hat a_1, \hat a_2, ... \hat a_M$) are the spatially localized rectangular modes of width $\delta$, $j$-th mode center is placed at the position $x_j=j \delta$ in the object plane (see Figure \ref{fig:losses}.b).
 The one-photon input states corresponding to these modes, $\left\{ \ket{\psi_j} = \hat a_j^\dagger \ket{0} \right\}_{j \in \{1,2,...,M\}}$, span a Hilbert space $\mathcal{H}_\textrm{obj}$. States $\hat a_{M+1}^\dagger \ket{0}, ..., \hat a_{N}^\dagger \ket{0}$ span another Hilbert space $\mathcal{H}_\textrm{rest}$, orthogonal to $\mathcal{H}_\textrm{obj}$. The space $\mathcal{H}_\textrm{in} = \mathcal{H}_\textrm{obj} \oplus \mathcal{H}_\textrm{rest}$ contains all possible one-photon input states---some of them lie in the object area, another correspond to uncontrolled input modes. The space of all one-photon output states can be written as $\mathcal{H}_\textrm{out} = \mathcal{H}_\textrm{img} \oplus \mathcal{H}_\textrm{lost}$, where $\mathcal{H}_\textrm{img}$ contains states observable in the image plane, and $\mathcal{H}_\textrm{lost}$ contains photons in lost modes, which cannot be observed. Notice, that we do not assume anything about the origin of losses. The lost modes may correspond to photons physically absorbed by the aperture, but also to all photons that are not observed in the image plane (e.g. photons diffracted at a very large angle by the aperture). We assume, that our system performs a unitary transformation $\hat{U}$ between $\mathcal{H}_\textrm{in}$ and $\mathcal{H}_\textrm{out}$, which implies that $\textrm{dim} \mathcal{H}_\textrm{in} = \textrm{dim} \mathcal{H}_\textrm{out}$, but does not imply anything about $\textrm{dim} \mathcal{H}_\textrm{img}$ and $\textrm{dim} \mathcal{H}_\textrm{obj}$---the presence of arbitrarily large spaces $\mathcal{H}_\textrm{rest}, \mathcal{H}_\textrm{lost}$ makes our model very general.

Each state $\ket{\psi_j} \in \mathcal{H}_\textrm{obj}$ is transformed into an output state 
\begin{equation}
\label{phiout}
\ket{\phi_j} = \hat U \ket{\psi_j} = \sqrt{1-p_j}\ket{\phi_{j}^\textrm{lost}} + \sqrt{p_j} \ket{\phi_{j}^\textrm{img}},
\end{equation}
 where $\ket{\phi_{j}^\textrm{lost}} \in \mathcal{H}_\textrm{lost}$, $\ket{\phi_{j}^\textrm{img}} \in \mathcal{H}_\textrm{img}$, and $p_j$ is the probability that a photon from a $j$-th rectangular mode is transmitted through the system. The system is translationally invariant, which means that
 \begin{equation}
 \label{pequal}
 p_1=...=p_M=p, 
 \end{equation}
 and that states $\ket{\phi_{j}^\textrm{img}}$ can be generated by shifting the state $\ket{\phi_{1}^\textrm{img}}$ in a position representation by a factor $(i-1) \delta K$, where $K$ is the system magnification. In other words,
 \begin{equation}
     \braket{x|\phi_{j}^\textrm{img}} = h(x - K \delta j),
 \end{equation}
 where $h$ is the system normalized cPSF provided that $\delta \rightarrow 0$.
 From unitarity of $U$ and orthonormality of $\{ \ket{\psi_j}\}$, we have
 \begin{equation}
 \label{djm}
     \delta_{jm} =
     \braket{\phi_j|\phi_m} =
     (1-p) \braket{\phi_{j}^\textrm{lost}|\phi_{m}^\textrm{lost}} + p \braket{\phi_{j}^\textrm{img}|\phi_{m}^\textrm{img}},
 \end{equation}
where we also used the fact that $\mathcal{H}_\textrm{img}$ is orthogonal to $\mathcal{H}_\textrm{lost}$. 
After summing \eqref{djm} over indices $j$ and $m$, we obtain
\begin{equation}
    M = (1-p) \braket{\Phi^\textrm{lost}|\Phi^\textrm{lost}} + p \braket{\Phi^\textrm{img}|\Phi^\textrm{img}},
\end{equation}
where $\ket{\Phi^\textrm{(img/lost)}} = \sum \ket{\phi_{j}^\textrm{(img/lost)}}$. We do not assume anything about $\ket{\Phi^\textrm{lost}}$, but its norm must be positive, which leads to the following upper-bound on $p$:
\begin{equation}
\label{pbound1}
    p \le \frac{M}{\braket{\Phi^\textrm{img}|\Phi^\textrm{img}}} = \frac{M}{\sum_{j,m} \braket{\phi_j^\textrm{img}| \phi_m^\textrm{img}}}.
\end{equation}
The knowledge of the normalized cPSF of the system allows to compute the right-hand side of the above inequality. Notice that in order to get the tightest bound for $p$, we can independently change the total size of the considered area $\Delta = \delta M$, and the size of a single mode $\delta$. When the derivative of $h(x)$ is bounded, then for any $\epsilon>0$ it is possible to choose $\Delta$ such that
\begin{equation}
\label{cond}
    \forall_{\Delta ' \in [-\Delta, \Delta]}\left| 1 - \int h(x) h^\ast (x- K \Delta') \textrm{dx} \right| \le \epsilon.
\end{equation}
After fixing $\Delta$, we can take the limit $\delta \rightarrow 0$, $M \rightarrow \infty$---then scalar products in the sum in the demominator of the right-hand side of \eqref{pbound1} become scalar products of the normalized cPSFs, and after using~\eqref{cond} and a triangle inequality, we obtain the following bound for $p$:
\begin{equation}
\label{pdelta}
    p \le \frac{1}{M (1-\epsilon)} = \frac{\delta}{\Delta(1-\epsilon)}.
\end{equation}
It is now clear, that for a single, small source of width $\delta$, placed in the object plane of our general imaging system, the ratio of transmitted signal goes to $0$ linearly with $\delta$. The constant in our bound for $p$ depends on the shape of normalized cPSF, and on our choice of $\epsilon$ and $\Delta$. For example, for a Gaussian PSF of width $\sigma$ and cPSF with a constant phase, we obtain
\begin{equation}
    p \le \frac{e^{1/2}}{2} K \frac{\delta}{\sigma},
\end{equation}
for $\epsilon = 1- e^{-1/2}$ and $\Delta=2 \sigma/K$ (this choice results in the tightest bound).

For a given value of $p$ and an arbitrary pure input state,
\begin{equation}
\label{pureinput}
    \ket{\psi} = \sum_{j=1}^M \alpha_j \ket{\psi_j},
\end{equation}
the probability of measuring a photon in one of the observed modes is the squared norm of the observable part of the output states, namely
\begin{equation}
\label{Pgen}
    T = p \sum_{j,m=1}^M \alpha^*_j \alpha_m \braket{\phi_{j}^\textrm{img}|\phi_{m}^\textrm{img}}.
\end{equation}
Let us now assume, motivated by $\eqref{pdelta}$, that $p= A \delta$, and compute $T$ in a limit $\delta \rightarrow 0$. In such a limit, each $\alpha_i$ should be proportional to $\sqrt{\delta}$ because of the normalization condition for the state $\ket{\psi}$. Therefore, we can write $\alpha_i =\psi_\textrm{in}(x_i) \sqrt{\delta} $ where $\psi_\textrm{in}(x)$ can be interpreted as the input one-photon wave function. Moreover,
\begin{equation}
    \braket{\phi_{j}^\textrm{img}|\phi_{m}^\textrm{img}} = \int h^*(x-K x_j) h(x-K x_m)\textrm{d}x,
\end{equation}
summation over $j$ and $m$ in \eqref{Pgen} can be replaced with integration over $x_j$ and $x_m$, and then we obtain
\begin{equation}
\label{Tgeneral}
    T = 
    A K \int \left| \hat{\psi_\textrm{in}}  ( K \cdot k) \hat h(k) \right|^2 \textrm{d}k,
\end{equation}
where convolution theorem and Parseval's theorem were used. We see that the transmission ratio is of the same form as that obtained for the $4f$ system (up to a multiplicative constant $A$ and a magnification factor~$K$), and $A K \hat h(k)$ can be interpreted as the attenuation factor of a  spatial frequency $K \cdot k$.  It shows, that the results obtained at the beginning of this section for the $4f$ system, can be generalized to any passive, linear, translationally invariant optical imaging system.

\section{Partially coherent imaging}
\label{sec:coherence}
One of the main purposes of this paper is to resolve recent disputes concerning the influence of partial coherence between two small sources on the QFI associated with the estimation of their separation. The debate was started in \cite{larson2018resurgence}, where authors compute the QFI for partially coherent sources and claim, that even a small degree of coherence leads to the resurgence of the Rayleigh's curse. Tsang and Nair replied \cite{tsang2019resurgence} by showing that the SPADE measurement, which is optimal for incoherent case \cite{tsang, tsang2017subdiffraction}, allows to obtain classical FI larger than the QFI predicted in \cite{larson2018resurgence}, which is obviously a contradiction. Different approaches, resulting in contradictory results, were presented in a further discussion \cite{larson2019resurgence, hradil2019quantum, lee2019surpassing, tsang2019resolving, liang2020coherence, wadood2021superresolution}. It is assumed in \cite{larson2018resurgence, larson2019resurgence, liang2020coherence}, that the total signal power in the image plane depends neither on the sources separation nor on their degree of coherence, and therefore it is enough to consider single-photon states to obtain the total value of QFI. At the same time, the model used in \cite{tsang2019resurgence, lee2019surpassing} to compute classical FI for SPADE and direct imaging implicitly assumes photon losses, which depend on object parameters. The latter approach is also supported by experimental work \cite{wadood2021superresolution}, but is not supplemented with the computation of the QFI, which leaves the question about the optimality of SPADE measurement open. Another approach is presented in \cite{hradil2019quantum}, where the cost of creating a coherent state from the purification of an incoherent one is studied in order to model photon losses. These results, undoubtedly very interesting from a theoretical point of view, do not necessarily correspond to real optical imaging systems, as they are inconsistent with  \cite{tsang2019resurgence, lee2019surpassing, wadood2021superresolution} and with our results from Section \ref{sec:losses}. 

The arguments presented in Section \ref{sec:losses} clearly show, that lossless models considered in \cite{larson2018resurgence, larson2019resurgence, liang2020coherence} are not physical. Moreover, in order to obtain a result applicable to any passive, linear, translationally invariant optical imaging system with a given normalized cPSF, it is enough to do all calculations for a 4f system with an appropriate aperture. 

Let us therefore consider a binary object, placed in plane (I) of a 4f system of magnification $1$ (see Figure~\ref{4f}). The object consists of two identical, small rectangular sources of width $\delta$ (the limit $\delta \rightarrow 0$ will be considered in order to examine point sources), separated by the distance $s \gg \delta$, with the centroid at $x=0$. The sources are weak, all approximations which were made to derive \eqref{eq5} are valid. The degree of coherence between two sources is $\gamma = r e^{i \varphi}$. When $\left| \gamma \right| < 1$, the object is not fully coherent, and \eqref{eq5} cannot be used directly---the proper description of such an object requires the use of the coherent mode decomposition \cite{wolf1981new}. For the binary source considered, it is enough to use two orthogonal, one-photon modes (symmetric and anti-symmetric):
\begin{equation}
\label{modespm}
    \ket{\psi_\pm} = \frac{1}{\sqrt{2 \delta}} \int \left[ \textrm{rect}\left( \frac{x + s/2}{\delta} \right) \pm \textrm{rect}\left( \frac{x - s/2}{\delta} \right) \right]  \ket{x} \textrm{d} x.
\end{equation}
Provided that two sources are equally bright, the most general form of the density matrix describing the input field is
\begin{equation}
\label{rhoI24}
    \rho_\textrm{I} = p_\textrm{em} \rho_\gamma + \left(1 - p_\textrm{em} \right) \ket{0} \bra{0},
\end{equation}
where  
\begin{equation}
\label{rhog}
    \rho_\gamma = \frac{1}{2} \begin{bmatrix} 1 + \textrm{Re}(\gamma) & i \textrm{Im} (\gamma) \\ -i \textrm{Im} (\gamma) & 1 - \textrm{Re}(\gamma) \end{bmatrix}
\end{equation}
in the basis $\ket{\psi_\pm}$. Factor $\gamma$ in \eqref{rhog}  must satisfy $\left| \gamma \right|\le 1$, and can be interpreted as the degree of coherence betweeen the sources. 

 For simplicity, let us assume that the aperture is symmetric, $A(x)=A(-x)$. The transformation between planes (III) and (IV), performed by an ideal lens, is unitary, and does not affect the QFI associated with $s$ estimation. It is therefore possible to compute the QFI for the state in plane (III), and then interpret it as the QFI connected to the actually observed field. The density matrix of plane (III) can be obtained by transforming $\rho_\textrm{I}$ usign transformation rules similar to those described in \eqref{sout1} and \eqref{sout2} --- the only difference is that $\phi_\textrm{out}(x)$ from \eqref{sout2} should be replaced with its inverse Fourier transform.
 When $\gamma = e^{i \varphi}$, the wave function of the one-photon part of a state in plane~(III) is
\begin{equation}
\label{xphi26}
    \braket{x| \phi_\varphi} = \mathcal{N}(\varphi,s) A(x) ~ \textrm{sinc}\left( \pi x \delta \right) \cos \left(\pi x s + \phi/2 \right),
\end{equation}
where $\mathcal{N}(\varphi,s)$ is a normalization factor. It is useful to define a family of non-normalized states $\ket{\tilde \phi_\varphi} = \mathcal{N}^{-1} \ket{\phi_\varphi}$.
The above definition allows us to write $\rho_\textrm{III}$ for a general $\gamma$ as
\begin{equation}
\label{rhoIII}
    \rho_\textrm{III} = 2 \delta p_\textrm{em} \sum_{i,j=0}^1 \rho_\gamma^{(ij)} \ket{\tilde \phi_{i \pi}} \bra{\tilde \phi_{j \pi}} + p_0 \ket{0} \bra{0},
\end{equation}
 where $p_0$ is the probability of a zero-photon state (see Appendix \ref{apA} for the details of the computations). The system transmission for a general, partially coherent input state is the trace of the one-photon part of $\rho_\textrm{III}$ divided by $p_\textrm{em}$, which simplifies to
 \begin{equation}
 \label{T}
     T(\chi,s) = 2 \delta \braket{\tilde \phi_\chi|\tilde \phi_\chi}
 \end{equation}
where $\cos \chi = \textrm{Re}(\gamma)$. The parameter $\chi$ is useful because in many situations a partially coherent state with the degree of coherence $\gamma$ is equivalent to a fully coherent state with the degree of coherence $e^{i \chi}$.

We are now ready to derive the fundamental bound on the precision of $s$ estimation. The general formula for the QFI associated with the quantum states family $\rho(s)$ is
 \begin{equation}
 \label{QFIdef}
     \mathcal{F}[\rho(s)] = \textrm{Tr}( \rho \Lambda^2),
 \end{equation}
 where $\Lambda$ matrix is defined by the equation
 \begin{equation}
 \label{Lambdadef}
     \frac{\textrm{d} \rho}{\textrm{d} s} = \frac{1}{2} \left( \Lambda \rho + \rho \Lambda  \right). 
 \end{equation}
 
 Coefficients $\rho_\gamma^{(ij)}$ in \eqref{rhoIII} do not depend on $s$, which, supplemented by some orthogonality relations between non-normalized states $\ket{\tilde{\phi}_{i \pi}}$ and their derivatives (specified in Appendix \ref{apA}), allows us to provide a simple formula for the QFI associated with the observed state,
\begin{equation}
 \label{FQB}
     \mathcal F [\rho_\textrm{III}(s)] = 8 \delta p_\textrm{em}  \sum_{i =0}^1 \rho_\gamma^{(ii)} \braket{\dot{ \tilde{ \phi}}_{i \pi}| \dot{ \tilde{ \phi}}_{i \pi}} + \frac{1}{p_0} \left( \frac{\textrm{d} p_0}{ \textrm{d}s } \right)^2,
 \end{equation}
 where $\ket{\dot{ \tilde{ \phi}}_\varphi} \equiv \partial_s \ket{ \tilde{ \phi}_\varphi}$, and the last term is of the order of $\delta^2$, and can be neglected when $\delta \rightarrow 0$. 
  From \eqref{rhog} and \eqref{FQB} follows that the QFI does not depend on $\textrm{Im}(\gamma)$. Probability $p_\textrm{em}$ depends on an arbitrary observation time, we can get rid of this non-physical parameter by computing QFI per emitted photon, which simplifies to (see Appendix \ref{apA})
  \begin{equation}
  \label{Fem1}
      \mathcal{F}_\textrm{em} [\rho_\textrm{III}(s)] = \mathcal{F} [\rho_\textrm{III}(s)]/p_\textrm{em} = 8 \delta \braket{\dot{\tilde \phi}_\chi|\dot{\tilde \phi}_\chi} + \mathcal{O}(\delta^2).
  \end{equation}
   $\mathcal{F}_\textrm{em}$ drops down to $0$ for $\delta \rightarrow 0$, as all photons are absorbed by the aperture when sources are infinitesimally small, see also \eqref{T}. To obtain the quantity which does not vanish for point sources, one can compute the QFI per detected photon,
  \begin{equation}
  \label{Fdetgen}
      \mathcal{F}_\textrm{det}  = \frac{\mathcal{F}_\textrm{em}}{T (\chi,s )}.
  \end{equation}
  \begin{figure*}[t]
    \centering
    \includegraphics[width=1.\textwidth]{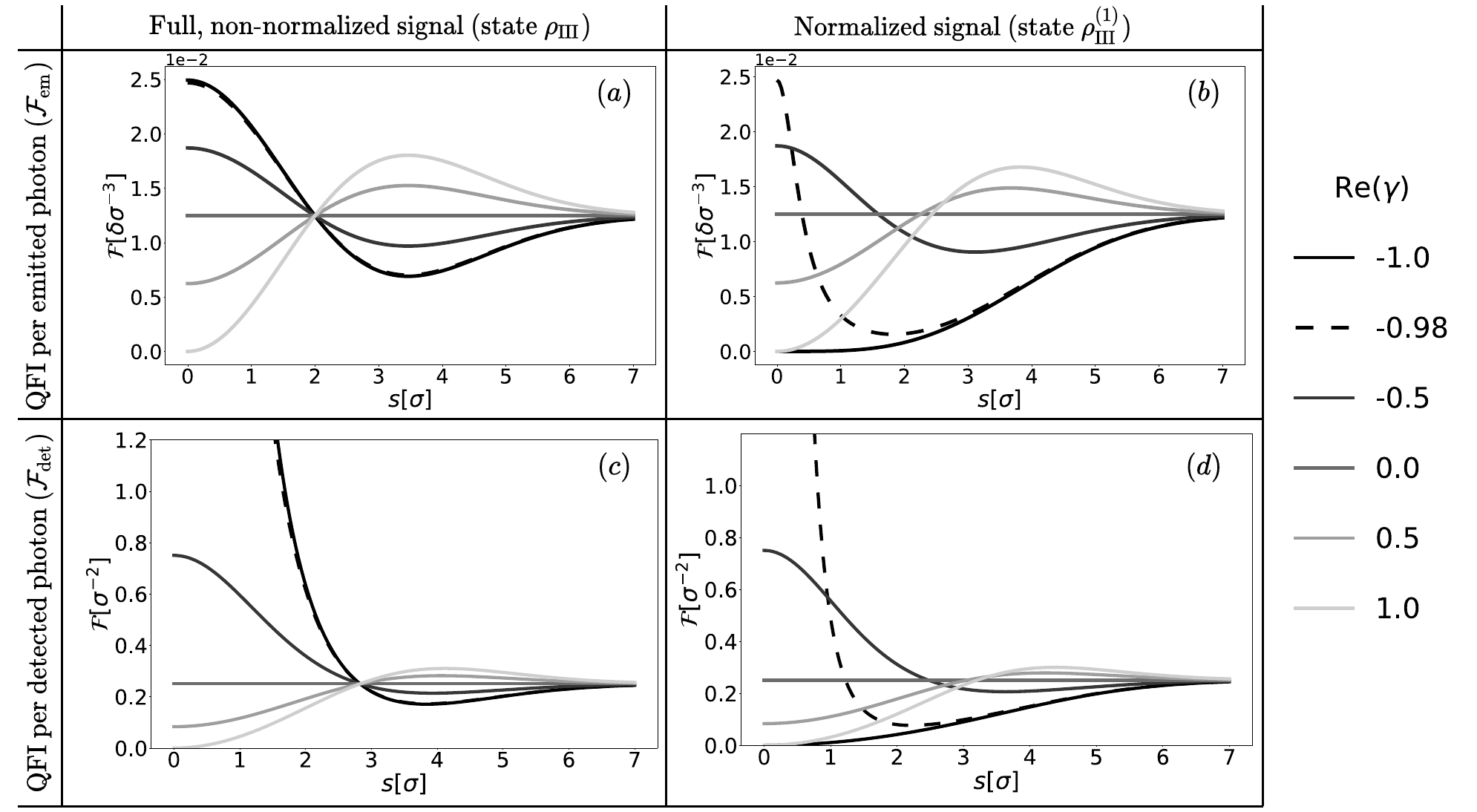}
    \caption{The QFI for Gaussian aperture, $A(x) = \exp \left( - \frac{4 \pi^2 \sigma^2 x^2}{\lambda^2 f^2} \right)$, which leads to a Gaussian PSF of width $\sigma$, is shown for different values of $\textrm{Re}(\gamma)$. Full quantum state ($\rho_\textrm{III}$), and its normalized single-photon part ($\rho_\textrm{III}^{(1)}$) are considered, two possible normalization conventions are shown. Part (a) is consistent with classical FI computed in \cite{tsang2019resurgence}, whereas part (d) is consistent with the QFI obtained in \cite{larson2019resurgence,hradil2019quantum}. Notice the substantial difference between $\textrm{Re}(\gamma)=-1$ and $\textrm{Re}(\gamma)=-0.98$ for cases (b) and (d), and the difference of dimensions on $y$ axis between $\mathcal{F}_\textrm{det}$ and $\mathcal{F}_\textrm{em}$.}
    \label{4plots}
 \end{figure*}
 However, it is better physically justified to treat the number of emitted photons as a resource, as, for example, it is directly connected with the measurement time. 
  
  Let us see what happens if the state $\rho_\textrm{III}$, defined in \eqref{rhoIII}, is replaced with its normalized one-photon part,
  \begin{equation}
  \label{rhop1}
      \rho_\textrm{III}^{(1)} = \braket{\tilde \phi_\chi|\tilde \phi_\chi}^{-1} \sum_{i,j=0}^1 \rho_\gamma^{(ij)} \ket{\tilde \phi_{i \pi}} \bra{\tilde \phi_{j \pi}}
  \end{equation}
  The QFI computed for this single-photon state is already QFI per detected photon, and is equal to (see Appendix \ref{apA} for the derivation)
  \begin{equation}
  \label{Fdetrho1}
      \mathcal{F}_\textrm{det}[\rho_\textrm{III}^{(1)}(s)] = \mathcal{F}[\rho_\textrm{III}^{(1)}(s)] = 4 \left( \braket{\dot \phi_\chi | \dot \phi_\chi} - |\braket{\dot \phi_\chi|\phi_\chi}|^2 \right).
  \end{equation}
   
 When $|\gamma|=1$, then $\rho_\textrm{III}^{(1)}=\ket{\phi_\chi}\bra{\phi_\chi}$, and \eqref{Fdetrho1} is just a well-known formula for QFI for a pure states family $\ket{\phi_\chi}$. The QFI per emitted photon, $\mathcal{F}_\textrm{em}[\rho_\textrm{III}^{(1)}]$, obtained with the help of \eqref{Fdetgen}, is smaller than $\mathcal{F}_\textrm{em}[\rho_\textrm{III}]$, as the following relation holds:
 \begin{equation}
 \label{Femrho1}
 \mathcal{F}_\textrm{em}[\rho_\textrm{III}^{(1)}(s)] = \mathcal{F}_\textrm{em}[\rho_\textrm{III}(s)] - 8 \delta |\braket{ \phi_\chi|\dot{\tilde \phi}_\chi}|^2.
 \end{equation}
 This reflects the fact, that some information about $s$ is encoded in the total power of observed signal.

 To conclude, we obtained four different results---QFI per detected or emitted photon, for states $\rho_\textrm{III}$ and $\rho_\textrm{III}^{(1)}$. It is more natural to use QFI per emitted photon, but the problem with this quantity is that it vanishes for $\delta \rightarrow 0$. The QFI for $\rho_\textrm{III}^{(1)}$ was computed in \cite{larson2018resurgence,larson2019resurgence,liang2020coherence}, but in the light of presented arguments, it is clear that the real physical situation is properly described by the state $\rho_\textrm{III}$. However, in many practical situations, the power of light emitted by the object is not known, and then it is problematic to extract any information from the total power of the observed signal---in such a case, the QFI associated with $\rho_\textrm{III}^{(1)}$ can serve as a reasonable upper-bound for the precision of estimation of $s$.
 
 We studied the dependence of the QFI on $s$ and $\textrm{Re}(\gamma)$ for Gaussian aperture $A(x)$, which leads to a Gaussian PSF with standard deviation $\sigma$. The results are shown in Fig.~\ref{4plots} (see Appendix \ref{apB} for analytical expressions). One can observe, that curves shown in part (a) of Fig.~\ref{4plots}, which correspond to $\mathcal{F}_\textrm{em} [ \rho(s)]$, coincide with those presented in Fig. 1 in \cite{tsang2019resurgence}, where the classical FI for SPADE measurement is computed. Indeed, our analysis shows, that SPADE measurement remains optimal if partial coherence between sources arises---see Appendix \ref{apB} for analytical proof. By comparing the first and second column of Fig.~\ref{4plots}, we can figure out, how important the information hidden in the total signal power variability is---generally speaking, this source of information is crucial for $\textrm{Re}(\gamma)<0$, and that is the reason why the discrepancies in the previous works were manifested mainly for negative degrees of coherence. Notice that the Rayleigh curse is really inevitable only for $\gamma=1$, whereas for $\gamma=-1$, it can be eluded if the full, non-normalized signal is properly used. What may surprise the reader, is that $\mathcal{F}_\textrm{em}$ for a fixed $s / \sigma$ is proportional to $\delta \sigma^{-3}$, not to $\sigma^{-2}$ as in \cite{tsang,tsang2019resurgence} and other references. This discrepancy is caused by the fact, that the number of photons detected from a single source depends on the aperture size, and consequently on $\sigma$. This effect is often hidden in a normalization factor, which in fact depends on $\sigma$ (e.g. factor $N$ in \cite{tsang}, factor $N_0$ in \cite{tsang2019resurgence}).
 \section{The issue of vanishing transmission}
So far, we tried to get around the fact, that no signal is observed in the image plane when the object consists of really point sources, and the imaging system is translationally invariant. However, this vanishing transmission may be slightly worrisome because the mentioned assumptions are omnipresent in the theoretical works about optical imaging. 
It is worth noting that in the derivation of \eqref{pdelta}, we did not even assume full-plane translational invariance (we did this later to derive \eqref{Tgeneral})---in fact, \eqref{pdelta} is valid, when the system is translationally invariant in some arbitrarily small finite region. In order to construct a model, in which the observed signal does not vanish, one needs to either consider finite-size, internally coherent sources, or relax the translationally invariance assumption. The first approach seems reasonable in microscopy, whereas the latter one should be more suitable for astronomy, where the size of emitters is neglible compared to the lenghtscale at which it possible to observe any structure of the object. 

\begin{figure}[t]
\centering
\includegraphics[width=1. \linewidth]{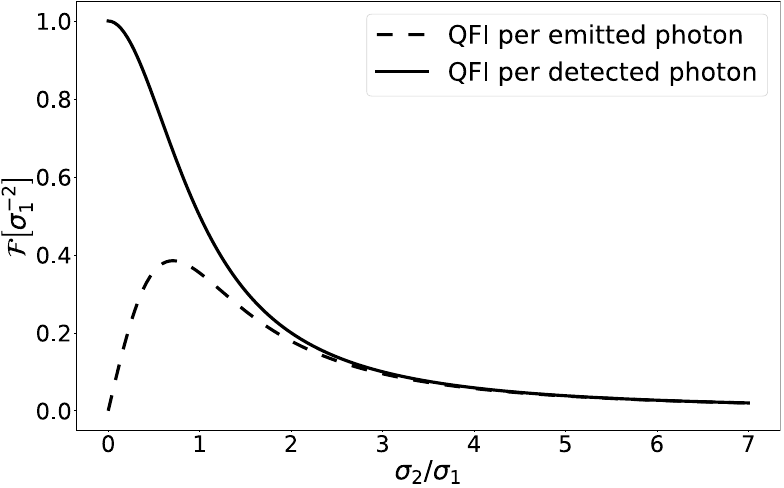}
\caption{ The QFI associated with the localization of an internally coherent, Gaussian object of width $\sigma_2$. The image of an object is created using a translationally invariant system with the Gaussian PSF of width $\sigma_1$. }
\label{fig:gaussl}
\end{figure}

Both modifications may affect the fundamental bounds obtained in the basic model. Let us consider, as an example, the problem of a single light source localization. When a point source is imaged by a translationally invariant system with a Gaussian PSF of width $\sigma_1$, then the QFI per detected photon, associated with the estimation of its position is \cite{helstrom1970estimation,Tsang:15}
\begin{equation}
\label{onelimit}
    \mathcal{F}_\textrm{det} =  \sigma_1^{-2}.
\end{equation}
Problems start to appear, when one wants to compute a better physically motivated QFI per emitted photon, $\mathcal{F}_\textrm{em}$--- the transmission in the considered case is $0$, so $\mathcal{F}_\textrm{em} = 0$. In order to fix this issue, let us change our model, and assume, that the source to be localized is described by a Gaussian wave function
\begin{equation}
    \psi_\textrm{in}(x) = \frac{1}{\sqrt[4]{2 \pi \sigma_2^2} } \textrm{exp} \left( - \frac{(x-x_0)^2}{4 \sigma_2^2} \right),
\end{equation}
and our task is to estimate the position of its center~$x_0$. The imaging process can be modeled with the help of a 4f system, with the Gaussian aperture, like in Section \ref{sec:coherence}. After using \eqref{sout1}, we obtain the expression for the output state, $\rho_\textrm{out} (x_0)$, in which the information about losses is encoded. This allows us to obtain the following results
\begin{equation}
    \mathcal{F}_\textrm{em} \left[ \rho_\textrm{out} \left( x_0 \right) \right] = \frac{1}{ \sigma_1^2}\frac{\sigma_2/\sigma_1}{\left(1+(\sigma_2/\sigma_1)^2 \right)^{3/2}},
\end{equation}
\begin{equation}
    \mathcal{F}_\textrm{det} \left[ \rho_\textrm{out} \left( x_0 \right) \right] = \frac{1}{ \sigma_1^2} \frac{1}{1+(\sigma_2/\sigma_1)^2},
\end{equation}
which we present in Fig. \ref{fig:gaussl}.
As we can see, $\mathcal{F}_\textrm{det}$ vanishes both for $\sigma_2 \rightarrow 0$ (because $T \rightarrow 0$), and for $\sigma_2 \rightarrow \infty$ (because wider objects are harder to localize). The precision of the localization is the largest for $\sigma_2 = \sigma_1/\sqrt{2}$, and a well-known limit from \eqref{onelimit} can never be achieved, when a non-unit transmission is taken into account. One can expect, that fundamental limits associated with the imaging of objects consisting of many points \cite{tsang, Zhou2019} will be affected in a similar way after a necessary modification of the shape of single light sources, which leads to non-zero transmission.

The problem of vanishing transmission arises also in all models, in which the object plane is fully incoherent. Let us go back to the considerations from Section \ref{sec:losses}, and replace the general pure input state from \eqref{pureinput} with a mixed state of the form
\begin{equation}
    \rho^\textrm{in} = \sum_{j=1}^M \left| \alpha_j \right|^2 \ket{\psi_j}\bra{\psi_j}.
\end{equation}
After applying a unitary evolution $\hat U$ defined in \eqref{phiout} to $\rho_\textrm{in}$, and using \eqref{pequal}, we obtain
\begin{equation}
    \rho^\textrm{out} = \hat U \rho^\textrm{in} \hat U^\dagger = p \rho^\textrm{img} + (1-p) \rho^\textrm{lost},
\end{equation}
where
\begin{equation}
    \rho^\textrm{img/lost} = \sum_{j=1}^M \left| \alpha_j \right|^2 \ket{\phi_j^\textrm{img/lost}} \bra{\phi_j^\textrm{img/lost}}.
\end{equation}
are normalized states corresponding to observed and lost photons respectively.
It is therefore clear, that the transmission is always equal to $p$ for a diagonal $\rho_\textrm{in}$. Moreover, the size of a single mode ($\delta$) is the maximal distance between two mutually coherent parts of the object. In order to describe a fully incoherent object, we should take the limit $\delta \rightarrow 0$. But then, according to \eqref{pdelta}, $p \rightarrow 0$, which means that the transmission vanishes.

Recently, the problem of imaging of a continuous, fully incoherent, 1D object characterized by intensity distribution $I(x)$, has been rigorously studied using quantum estimation-theory tools \cite{semiparametric, incoherentc1, incoherentc2}. As a result, expressions for the QFI associated with the estimation of spatial moments of $I(x)$ have been derived. However, it is now clear, that the model considered in the mentioned works result in zero transmission. In order to obtain physically relevant results, one should assume that the spatial coherence in the object plane is present at sufficiently small length scales (or consider a non-translationally invariant imaging system). Further studies are required to figure out, how fundamental losses affect the precision limits in super-resolution imaging of complex objects.
\section{Summary}
We have shown, that the fundamental role of photon losses has been underestimated in many recent theoretical works about super-resolution imaging. As we demonstrated, losses cannot be omitted even in the most idealized and simplified considerations, because lossless models are in contradiction with the principles of quantum mechanics.
In order to clarify the issues associated with the behavior of the QFI for the separation between two partially coherent sources, we have carefully studied the imaging process using the well-established $4f $ system-based model. We have satisfactorily explained the discrepancies in previous works that were caused by an unphysical assumption, that the transmission of an imaging apparatus with non-trivial PSF is $100\%$ for all input objects.  Properly computed QFI coincides with the previously calculated classical FI for SPADE measurement \cite{tsang2019resurgence}, which indicates the optimality of this measurement. 

We have also shown that some fundamental limitations in super-resolution imaging are questionable as they were derived from non-physical models leading to zero transmission. There are different ways to construct physical models with a non-vanishing transmission, but all of them will probably lead to some complications, especially in the case of the imaging of complex, continuous objects.

\textit{References \cite{tsang2021quantum} and \cite{wadood2021superresolution} are independent works. In \cite{tsang2021quantum}, Tsang provides a more general formula for the QFI, which can be applied for partially coherent sources, and properly takes photon losses into account.  In this paper, we provide different physical arguments supporting the same thesis, and study the specific case of weak, partially coherent sources in more detail. Experimental results presented in \cite{wadood2021superresolution} agree with the theoretical analysis presented in Section \ref{sec:coherence}. The total signal power variations as a source of information are also studied in another parallel work \cite{hradil2021exploring}. However, as we discussed above, the connection between the model proposed by the authors and the performance of a realistic imaging system is not clear---in particular, the losses associated with $\gamma=0$ should not be neglected, as they are higher than e.g. losses for $\gamma=1$. }

\label{sec:point}
 \section*{Acknowledgments}
 First of all, I would like to thank Rafał Demkowicz-Dobrzański for very fruitful discussions. I also thank Konrad Banaszek for sharing his experience and valuable remarks, and Francesco Albarelli for his comments and suggestions concerning the latest literature. This work is a part of the project ''Quantum  Optical  Technologies''  carried  out  within the   International Research Agendas programme of the Foundation for Polish Science cofinanced by the European Union under the European Regional Development Fund.

\bibliographystyle{quantum} 
\bibliography{sample}

\newpage
\onecolumngrid
\appendix

\section{The details of QFI calculations}
\label{apA}
In this section the details concerning the derivation of Eqs. \eqref{rhoIII}, \eqref{T}, \eqref{FQB}, \eqref{Fem1}, \eqref{Fdetrho1}, \eqref{Femrho1} from the main text are provided. 

Let us start with defining the following family of one-photon input states:
\begin{equation}
\label{27}
    \ket{\psi_\varphi} = \frac{1}{\sqrt{2 \delta}} \int \left[ e^{i \varphi/2} \textrm{rect}\left( \frac{x + s/2}{\delta} \right) + e^{-i \varphi/2} \textrm{rect}\left( \frac{x - s/2}{\delta} \right) \right]  \ket{x} \textrm{d} x.
\end{equation}
Notice that $\ket{\psi_0} = \ket{\psi_+}$, $\ket{\psi_\pi} = \ket{\psi_-}$ up to a global phase, if we refer to the definition from \eqref{modespm} from the main text.
In order to obtain the equation for $\rho_\textrm{III}$ \eqref{rhoIII}, we should start from \eqref{rhoI24}, and transform the state $\rho_\textrm{I}$ linearly, according to the following rule
\begin{equation}
\label{trans}
    \ket{\psi_{\varphi}} \bra{\psi_{\varphi}} \rightarrow T(\varphi,s) \ket{\phi_\varphi} \bra{\phi_\varphi} + \left(1-T(\varphi,s) \right) \ket{0} \bra{0},
\end{equation}
where $\ket{\phi_\varphi}$ is defined in \eqref{xphi26}, and $T(\varphi, s)$ can be calculated with the help of (9):
\begin{equation}
    T(\varphi, s) = \int_{-\infty}^{\infty} \left| \hat \psi_\varphi(x) A(x) \right|^2 = 2 \delta \int_{-\infty}^{\infty} |A(x)|^2 ~ \textrm{sinc}^2( \pi x \delta) \cos^2(\pi x s + \phi/2) \textrm{d}x.
\end{equation}
The explicit formula for the non-normalized state $\ket{ \tilde \phi_\varphi}$ is
\begin{equation}
\label{appsi}
    \ket{\tilde \phi_\varphi} = \int_{-\infty}^{\infty} A(x) ~ \textrm{sinc}(\pi x \delta) \cos(\pi x s + \varphi/2) \ket{x} \textrm{d}x.
\end{equation}
From normalization condition, $\braket{\phi_\varphi|\phi_\varphi} = |\mathcal{N}(\varphi,s)|^2 \braket{\tilde \phi_\varphi| \tilde \phi_\varphi}=1$, the normalization constant satisfies
\begin{equation}
\label{apc}
    |\mathcal{N}(\varphi,s)|^{-2} = \int_{-\infty}^{\infty} |A(x)|^2 ~ \textrm{sinc}^2( \pi x \delta) \cos^2(\pi x s + \phi/2) \textrm{d}x = \frac{T(\varphi, s)}{2 \delta},
\end{equation}
which means that
\begin{equation}
\label{Tphi}
    T(\varphi, s) \ket{\phi_\varphi} \bra{\phi_\varphi} = 2 \delta \ket{\tilde \phi_\varphi} \bra{\tilde \phi_\varphi}.
\end{equation}
After inserting \eqref{Tphi} into the transformation (\ref{trans}), and applying it to $\rho_\textrm{I}$, we obtain the formula for $\rho_\textrm{III}$ \eqref{rhoIII}. 
The transmission of the system for an arbitrary partially coherent input is
\begin{equation}
\label{Tgamma}
    T = 2 \delta \textrm{Tr} \left[ \sum_{i,j=0}^1 \rho_\gamma^{(ij)} \ket{\tilde \phi_{i \pi}} \bra{\tilde \phi_{j \pi}} \right] =  \delta \left[ (1+\textrm{Re}(\gamma)) \braket{\tilde \phi_0|\tilde \phi_0} + (1-\textrm{Re}(\gamma)) \braket{\tilde \phi_\pi|\tilde \phi_\pi} \right], 
\end{equation}
where we used orthogonality property $\braket{\tilde \phi_0 | \tilde \phi_\pi}=0$.
From \eqref{appsi} follows that
\begin{equation}
\label{A10}
    \ket{\tilde \phi_\varphi} = \cos \frac{\varphi}{2} \ket{\tilde \phi_0} + \sin \frac{\varphi}{2} \ket{\tilde \phi_\pi},
\end{equation}
which, after replacing $\textrm{Re}(\gamma)$ with $\cos \chi$, allows us to simplify \eqref{Tgamma} to the form
\begin{equation}
    T = T(\chi,s) = 2 \delta \braket{\tilde \phi_\chi| \tilde \phi_\chi},
\end{equation}
exactly as in \eqref{T}.

Let us now introduce the lemma, which simplifies the computations of the QFI significantly.

\textit{Lemma.---} Let us define the family of quantum states indexed by a real parameter $s$,  
\begin{equation}
\rho(s) = \sum_{i,j=1}^N B_{ij} \ket{\Psi_i(s)} \bra{\Psi_j(s)},
\end{equation}
where coefficients $B_{ij}$ do not depend on $s$, and vectors $\ket{\Psi_i}$ and their derivatives $\ket{\dot \Psi_i} \equiv \partial_s \ket{\Psi_i}$ satisfy
\begin{equation}
\label{assumptions}
    \braket{\Psi_i | \Psi_j} = \braket{\dot \Psi_i| \Psi_j} = \braket{\dot \Psi_i| \dot \Psi_j} = 0~~\textrm{for}~~i \ne j,
\end{equation}
and moreover
$
    \braket{\dot \Psi_i| \Psi_i} \in \mathbb{R}.
$
Then the QFI associated with the estimation of $s$ is
\begin{equation}
    \mathcal{F}\left[\rho(s) \right] = 4 \sum_{i=1}^N B_{ii} \braket{\dot \Psi_i| \dot \Psi_i}
\end{equation}

\textit{Proof.---} We are going to compute the QFI from its definition \eqref{QFIdef}. It is easy to check, that the following ansatz,
\begin{equation}
    \Lambda = 2 \sum_{i=1}^N \frac{\ket{\dot \Psi_i}\bra{\dot \Psi_i}}{\braket{\Psi_i|\dot \Psi_i}},
\end{equation}
makes the condition from \eqref{Lambdadef} satisfied. Indeed,
\begin{equation}
    \frac{1}{2} \left( \rho \Lambda + \Lambda \rho \right) = \sum_{i,j=1}^N B_{ij} \ket{\Psi_i} \bra{\Psi_j} \sum_{k=1}^N \frac{\ket{\dot \Psi_k}\bra{\dot \Psi_k}}{\braket{\Psi_k|\dot \Psi_k}} + \sum_{k=1}^N \frac{\ket{\dot \Psi_k}\bra{\dot \Psi_k}}{\braket{\dot \Psi_k| \Psi_k}} \sum_{i,j=1}^N B_{ij} \ket{\Psi_i} \bra{\Psi_j} = \sum_{i,j=1}^N B_{ij} \left( \ket{\dot \Psi_i}\bra{\Psi_j} + \ket{\Psi_i}\bra{\dot \Psi_j} \right),
\end{equation}
and the last expression is equal to $\frac{\textrm{d} \rho}{\textrm{d}s}$.
Furthermore, we have
\begin{equation}
    \Lambda^2 = 4 \sum_{i=1}^N \frac{\braket{\dot \Psi_i|\dot \Psi_i}}{(\braket{\Psi_i|\dot \Psi_i})^2} \ket{\dot \Psi_i}\bra{ \dot\Psi_i},
\end{equation}
and straightforward calculations show that
\begin{equation}
\label{thesis}
    \mathcal{F} = \textrm{Tr}(\rho \Lambda^2) = 4 \sum_{i=1}^N B_{ii} \braket{\dot \Psi_i| \dot \Psi_i},
\end{equation}
Q.E.D.


Let us write $\rho_\textrm{III}$ in a form
 \begin{equation}
     \rho_\textrm{III} = 2 \delta p_\textrm{em} \sum_{i,j=0}^1 \rho_\gamma^{(ij)} \ket{\Psi_i} \bra{\Psi_j} + \ket{\Psi_2} \bra{\Psi_2}
 \end{equation}
 where  $\ket{\Psi_0}= \ket{ \tilde \phi_0}$, $\ket{\Psi_1}= \ket{\tilde \phi_\pi}$, $\ket{\Psi_2} = \sqrt{p_0} \ket{0}$, $\rho_\gamma$ is defined in \eqref{rhog}. With the help of \eqref{appsi} and parity condition for $A(x)$, it is straightforward to check that the assumptions of the lemma, listed in \eqref{assumptions}, are satisfied for $\ket {\Psi_i}$ defined as above. Then, \eqref{FQB} follows directly from the thesis of our lemma, \eqref{thesis}. In order to obtain the formula for the QFI per emitted photon, we should divide the total QFI by the probability of a photon emission, $p_\textrm{em}$. Then, after using the time derivative of \eqref{A10} and replacing $\textrm{Re}(\gamma)$ with $\cos \chi$, we obtain \eqref{Fem1}. The probability of a photon detection for a given $\gamma$ is $p_\textrm{em} T(\chi, s)$, which means that the relation between the QFI per emitted and detected photon is given by \eqref{Fdetgen}.
 


The computation for $\mathcal F[\rho_\textrm{III}^{(1)}]$ are analogous to those for $\mathcal F[\rho_\textrm{III}]$. We start with writing down \eqref{rhop1} in a form
\begin{equation}
     \rho_\textrm{III}^{(1)} =  \sum_{i,j=0}^1 \rho_\gamma^{(ij)} \ket{ \Psi_i} \bra{\Psi_j},
 \end{equation}
 where $\ket{\Psi_0} =  \mathcal{N}(\chi, s)  \ket{\tilde \phi_0}$, $\ket{\Psi_1} = \mathcal{N} (\chi,s)  \ket{\tilde \phi_\pi}$, vectors $\ket{\Psi_i}$ satisfy the lemma assumptions from \eqref{assumptions}. Therefore, using \eqref{thesis}, we obtain
 \begin{equation}
     \mathcal{F} \left[ \rho_\textrm{III}^{(1)} (s) \right] =  4 \left[ (1 +\textrm{Re} (\gamma)) \braket{ \dot \Psi_0| \dot \Psi_0} + (1-\textrm{Re}(\gamma)) \braket{ \dot \Psi_1| \dot \Psi_1} \right],
 \end{equation}
which means, that the QFI depends on $\textrm{Re}(\gamma)$, but not on $\textrm{Im}(\gamma)$. Therefore, we can assume that $\gamma = e^{i \chi} $, compute the resulting QFI and generalize the obtained result for any $\gamma$ satisfying  $ \textrm{Re}(\gamma) = \cos \chi$. For $\gamma = e^{i \chi} $, the considered one-photon output state is pure,
\begin{equation}
    \rho_\textrm{III}^{(1)}(s) = \ket{\phi_\chi} \bra{\phi_\chi},
\end{equation}
and a well-known formula for the QFI associated with the family of pure states leads to Equation \eqref{Fdetrho1}, valid for any $\gamma$ with a real part $\cos \chi$. In order to obtain the QFI per emitted photon, we should use the general relation~\eqref{Fdetgen}, which, applied to \eqref{Fdetrho1}, leads to \eqref{Femrho1}. 
 \section{Analytical expressions for Gaussian aperture}
 \label{apB}
 In this section we provide analytical expressions for different types of QFI and for the classical FI associated with SPADE measurement for the case $A(x) = \exp \left( - \frac{4 \pi^2 \sigma^2 x^2}{\lambda^2 f^2} \right)$, which is studied in Fig. 3. Such an aperture leads to the Gaussian PSF with width $\sigma$. This specific form of $A(x)$ was inserted into  Eqs. \eqref{Fem1}, \eqref{Fdetgen}, \eqref{Fdetrho1}, \eqref{Femrho1}, and the following results were obtained:
 \begin{equation}
 \label{Femgaus}
     \mathcal{F}_\textrm{em}[\rho_\textrm{III}(s)]= \frac{\delta }{8 \sqrt{2} \pi^{1/2} \sigma^3}  \left( 1- \left(1-\frac{s^2}{4 \sigma^2}\right) \textrm{Re}(\gamma) e^{-\frac{s^2}{8 \sigma^2}}  \right) + \mathcal{O}(\delta^2)
 \end{equation}
 
 \begin{equation}
     \mathcal{F}_\textrm{det}[\rho_\textrm{III}(s)] = \frac{1}{4 \sigma^2} \left( 1 + \frac{ 2\textrm{Re}(\gamma) e^{-\frac{s^2}{8 \sigma^2}} \left(\frac{s^2}{8 \sigma^2}-1 \right)}{1+\textrm{Re}(\gamma) e^{-\frac{s^2}{8 \sigma^2}}} \right)+ \mathcal O (\delta)
 \end{equation}
 
 \begin{equation}
     \mathcal{F}_\textrm{em}[\rho^{(1)}_\textrm{III}(s)] = \frac{\delta}{8 \sqrt{2} \pi ^{1/2} \sigma^3} \frac{  1- \left(\textrm{Re}(\gamma) e^{-\frac{s^2}{8 \sigma^2}}-\frac{s^2}{4 \sigma^2}\right) \textrm{Re}(\gamma) e^{-\frac{s^2}{8 \sigma^2}}   }{
   1+\textrm{Re}(\gamma) e^{-\frac{s^2}{8 \sigma^2}} } + \mathcal O (\delta^2)
 \end{equation}
 
 \begin{equation}
     \mathcal{F}_\textrm{det}[\rho^{(1)}_\textrm{III}(s)]  = \frac{1}{4 \sigma^2} \frac{  1- \left(\textrm{Re}(\gamma) e^{-\frac{s^2}{8 \sigma^2}}-\frac{s^2}{4 \sigma^2}\right) \textrm{Re}(\gamma) e^{-\frac{s^2}{8 \sigma^2}}   }{\left(
   1+\textrm{Re}(\gamma) e^{-\frac{s^2}{8 \sigma^2}} \right)^2} + \mathcal O (\delta)
 \end{equation}
We want to show that $\mathcal{F}_\textrm{em}\left[ \rho_\textrm{III}(s) \right]$ is equal to the classical FI for SPADE measurement obtained in \cite{tsang2019resurgence}. To do so, let us find the analytical form for the expected photon number in $q$-th measured mode, $n_q$, which is defined in Eq.~(2) in \cite{tsang2019resurgence}. After adopting the notation from \cite{tsang2019resurgence}, and using Eq.~(1) from \cite{tsang2019resurgence}, we obtain
\begin{equation}
\label{nq}
    n_q = N_0 \left[ I_+^2 + I_-^2 + 2 \textrm{Re}(\gamma) I_+ I_- \right],
\end{equation}
where 
\begin{equation}
    I_\pm =  \left( \sqrt{ 2 \pi} \sigma \right)^{-1/2} \int_{-\infty}^{\infty} \exp \left[-(x \pm s/2)^2/(4 \sigma^2)  \right] \phi_q(x) \textrm{d}x,
\end{equation}
where $\phi_q(x)$ correspond to Hermite-Gaussian modes,
\begin{equation}
    \phi_q(x) = \left( \frac{1}{2 \pi \sigma^2} \right)^{1/4} \frac{1}{\sqrt{2^q q!}} H_q \left( \frac{x}{\sqrt 2 \sigma} \right) \exp \left( - \frac{x^2}{4 \sigma^2} \right).
\end{equation}
Hermite polynomials, $H_q(x)$, can be defined using generating function,
\begin{equation}
    \exp(2xt-t^2) = \sum_{q=0}^\infty H_q(x) \frac{t^q}{q!}.
\end{equation}
After multiplying both sides of the above equation by $\exp(-(x-b)^2)$, and integrating over $x$, one obtains
\begin{equation}
    \int_{-\infty}^\infty H_q(x) \exp \left(-(x-b)^2 \right) \textrm{d}x = \sqrt{\pi} (2b)^q,
\end{equation}
which can be used to simplify the expression for $I_\pm$ to the form
\begin{equation}
    I_\pm = \frac{\exp\left[-s^2/(32 \sigma^2)\right]}{\sqrt{q !}} \left( \mp \frac{s}{4 \sigma} \right)^q.
\end{equation}
We can now insert the last equation to \eqref{nq} in order to get the expression for $n_q$, which can subsequently be used to compute the FI associated with the measurement of photon number in $q$-th mode,
\begin{equation}
    F_q = \frac{1}{n_q} \left( \frac{\partial n_q}{\partial s}  \right)^2.
\end{equation}
The total FI associated with SPADE measurement reads
\begin{equation}
\label{Fspade}
    F = \sum_{q=0}^\infty F_q = \frac{N_0}{2 \sigma^2} \left( 1- \left(1-\frac{s^2}{4 \sigma^2}\right) \textrm{Re}(\gamma) e^{-\frac{s^2}{8 \sigma^2}}  \right)
\end{equation}
Equations (\ref{Femgaus}) and (\ref{Fspade}) are equivalent---the only difference is that the factor $N_0$, which depends on the size and the shape of the sources, is not specified in the latter case. The above calculations prove the optimality of SPADE measurement for Gaussian PSF for arbitrary degree of coherence between the sources.

\end{document}